\documentclass{sig-alternate}
\usepackage{algorithmic}

\begin{document}

\conferenceinfo{ExtremeCom '12,} {March 10-14, 2012, Z\"{u}rich, Switzerland.}
\CopyrightYear{2012}
\crdata{978-1-4503-1264-6/12/03}

\title{Cooperative Caching based on File Popularity Ranking in Delay Tolerant Networks}

\numberofauthors{2} 
\author{
\alignauthor
Tiance Wang\\
       \affaddr{Princeton University}\\
       \affaddr{Dept. of Electrical Engineering}\\
       \affaddr{Princeton, New Jersey 08540}\\
       \email{tiancew@princeton.edu}
\alignauthor
Pan Hui\\
       \affaddr{Deutsche Telekom Laboratories}\\
       \affaddr{Ernst-Reuter-Platz 7}\\
       \affaddr{10587 Berlin, Germany}\\
       \email{pan.hui@telekom.de}
\and
\alignauthor       
Sanjeev R. Kulkarni\\
       \affaddr{Princeton University}\\
       \affaddr{Dept. of Electrical Engineering}\\
       \affaddr{Princeton, New Jersey 08540}\\
       \email{kulkarni@princeton.edu}
\alignauthor
Paul Cuff\\
       \affaddr{Princeton University}\\
       \affaddr{Dept. of Electrical Engineering}\\
       \affaddr{Princeton, New Jersey 08540}\\
       \email{cuff@princeton.edu}       
}
\date{7 Jan, 2011}

\maketitle
\begin{abstract}
Increasing storage sizes and WiFi/Bluetooth capabilities of mobile devices have made them a good platform for opportunistic content sharing. 
In this work we propose a network model to study this in a setting with two characteristics: 1. delay tolerant; 2. lack of infrastructure. Mobile users generate requests and opportunistically download from other users they meet, via Bluetooth or WiFi. The difference in popularity of different web content induces a non-uniform request distribution, which is usually a Zipf's law distribution. We evaluate the performance of different caching schemes and derive the optimal scheme using convex optimization techniques. The optimal solution is found efficiently using a binary search method. It is shown that as the network mobility increases, the performance of the optimal scheme far exceeds the traditional caching scheme. To the best of our knowledge, our work is the first to consider popularity ranking in performance evaluation. 
\end{abstract}


\section{Introduction}
Mobile devices have seen significant growth in their storage and wireless connection capabilities. Due to the development of web-based services, caching replacement strategies for web proxies has been an active area of research. As a result, proxy caching is used to reduce network bandwidth usage, user delays and load on the origin servers. However, traditional caching strategies are not ideal for a mobile environment. Cooperative caching has been proposed as an effective means of exploiting the huge potential in the storage and connection capabilities of the large number of mobile devices \cite{zhuo2011social} \cite{zhuocontact} \cite{ioannidis2010distributed}. In this work, we consider a delay-tolerant content sharing network built over a network of mobile users and wireless access points, where the users download content opportunistically from each other via short-range communications (e.g. Bluetooth or WiFi). If the requested content is not found within the prescribed time, the users will download it through the more expensive 3G network. 

\vspace{1mm}

Considerable attention has been paid to such content sharing systems \cite{ioannidis2010distributed}\cite{zhuo2011social} \cite{zhuocontact}. However, to the best of our knowledge, no work has considered the popularity difference of different network content. It is well known that the popularity distribution of the network contents approximately follows Zipf's law (including YouTube, the famous video sharing website \cite{breslau1999web} \cite{cha2007tube}). According to Zipf's law, if among all files, a file is the $k$th most likely to be requested,  then the probability of request, $p_k$, approximately follows $p_k \propto k^{-\alpha}$, where $0 < \alpha < 1$. Intuitively, more popular files should be cached more frequently (i.e. have more replicas). We do not restrict the file popularity distribution to any particular type, but will focus on the Zipf-like distribution. 

\vspace{1mm}

To illustrate the need for opportunistic downloading, we can conceive a scenario where many users with similar interest are geographically located in an area for a relatively long duration of time, such as an arena or a tourist destination. We assume users have mobile devices with limited storage sizes. The requested content is delay tolerant in nature. 

\vspace{1mm}

In this paper, we establish a mathematical model of mobile content sharing network based on file popularity distribution, user mobility and delay tolerance. We derive the optimal cache allocation and devise a strategy to achieve the optimal allocation. The rest of the paper is organized as follows. In Section 2 we will review related work in the area of cooperative caching. The model is established in Section 3.1. In Section 3.2, we formulate the problem of finding the optimal cache allocation as a convex program and provide a binary search algorithm to find the optimal solution. We then extend our model to include contact duration limitation in Section 3.3. Section 4 contains simulation results that evaluate the performance of several caching strategies using miss rate as the metric. A simple pushing strategy is used to achieve a near-optimal cache allocation. It is shown that as the network mobility increases, the performance of the optimal caching strategy and the pushing strategy becomes indistinguishable. We discuss possible extensions of our methods and results in Section 5, before concluding in Section 6.

\section{Related Works}
Cooperative caching for mobile networks has been studied widely in recent years \cite{zhuo2011social} \cite{zhuocontact} \cite{ioannidis2010distributed}. \cite{zhuo2011social} considers the scenario where due to user mobility, the requested data may be too large to be transmitted within a single contact with a user who has the data in its local storage. Therefore a large file is divided into small packets. However, in our work we maintain the assumption that each file can be transmitted with a single contact since (1) peer-to-peer connection through WiFi is much faster than 3G network, and (2) a file divided into packets could be treated as several requests and each packet may have some value to the user. \cite{ioannidis2010distributed} studies cooperative caching for mobile networks. It applies a distributed caching replacement based on users' computed policy in the absence of a central authority and uses a voting mechanism for nodes to decide which content should be placed. In their model, the mobile users are divided into classes, so as to heterogenize users in terms of storage capacities, mobility, and interest in content. A user encounters another user with a Poisson rate determined by the classes these two users belong to. Also, a user encounters an access point with a certain Poisson rate. The merit of their model is that each user has its own rate of request on webpages and the algorithm is fully distributed, meaning that the action each user takes is based on its own knowledge only. 

The strategy to migrate traffic from the cellular networks to the free and fast device-to-device networks, so as to mitigate the overloaded cellular network is referred to as mobile data offloading \cite{li2011multiple}\cite{han2011mobile}. \cite{li2011multiple} formulated the problem of offloading data with various sizes and delay tolerances as a submodular function maximization problem. Our work is along the same lines in the sense that users try to rely on device-to-device transmission and avoid generating cellular network traffic as much as possible, thereby achieving data offloading.

It is helpful to find a unified mathematical model to capture the fundamental natures of node mobility. \cite{zheng2004recent} summarized several common methods of mobility modelling. The most widely used and studied model is the random waypoint model. A host randomly chooses a destination (waypoint) and moves toward the destination in a straight line, with a constant speed randomly selected in $[v_{min},v_{max}]$. After reaching the destination, a host pauses for a random period before moving to the next destination. A recent investigation on the pattern of individual mobility \cite{gonzalez2008understanding} reveals that individual mobility patterns are largly indistinguishable after correcting for differences in travel distances and the inherent anisotropy of each trajectory. 

\section{Model}
It is validated experimentally in \cite{gao2009multicasting}\cite{conan2007characterizing} that the time between two consecutive contacts of a pair of users (inter-contact time) follows an exponential distribution. We therefore model the contacts between nodes as a Poisson process. The contact pattern between users and wireless access points is also modelled as a Poisson process. Moreover, the following factors affect the performance of the caching strategy and are therefore addressed in our model:

\begin{enumerate}
	\item Cache capacity. When the cache capacity is unlimited, all contents on the network could be stored in the cache. However, this is unrealistic as the cache of a single user usually has very limited storage. The whole network has much more content than a single mobile device could store. We set the cache capacity as a parameter in our model.
	
	\item User mobility. There are many models attempting to simulate realistic human or vehicle behaviors \cite{hsu2007modeling} \cite{gonzalez2008understanding}. However, in our problem it is only of interest to know the encounter rate, or the frequency that users contact each other and the access points. The actually path along which a mobile user moves is irrelevant. This simplifies our model since we do not need to consider the difference in mobility patterns. 
	
	\item Density of access points. This is captured by the encounter rate $\lambda_{ap}$ of the users and the access points. The inter-contact time between a user and access points is different from the inter-contact time between mobile users. It follows an exponential distribution with a mean of $\lambda_{ap}$. 
	
	\item Distribution of requests. The distribution of requests directly affects the caching strategy. Intuitively, a more popular file should be made more accessible, and hence should be pushed into the network more frequently. Previous research indicates that web file requests follows a Zipf-like distribution \cite{breslau1999web} or truncated Zipf-like distribution, in which the request rate decays exponentially below a certain popularity level. 
	
	\item Contact duration. The contact duration also exhibits a Zipf-like distribution. If the contact duration is too short, a complete file could not be transmitted during a contact. For simplicity, we first assume that it only takes a single contact to complete the transmission. As a second step, we will consider the case where a single contact is not enough to complete the transmission.
\end{enumerate}

\subsection{First Model: No Contact Duration Limit}
We assume all users are statistically identical, which means they have the same cache capacity, mobility and request distribution. 
Considering the above factors, we now provide a formal problem statement.

	\vspace{5mm}

	\textbf{Problem statement}: The content distribution network hosts $N$ files, numbered $1,2,\dots,N$. The requests follow the distribution $P=\{p_1,\dots,p_N\}$ with $p_1+p_2+\dots+p_N=1$, i.e. when a request occurs, the probability that the request is for file $i$ equals $p_i$. Suppose the files are sorted by popularity. That is, $p_i>p_j$ for $i<j$. Each user generates requests independent of other users, and independent of previous requests. If the request results in a miss, then the mobile user downloads the file through the 3G network, which is slower and more expensive. Given a certain node mobility model and the location of wireless access points, we want to design a process of selecting and storing files in each mobile user's cache to maximize the probability that a user finds the requested file from a nearby user's cache (the \textbf{hit rate}).

\vspace{3mm}

	To simplify the analysis, we make the following assumptions:
	
	\begin{enumerate}
	\item All files are of equal size, and all users can store $K$ files in their cache.
	\item User encounters are modelled as independent Poisson processes with rate $\lambda$. User encounters with the wireless access points are also Poisson processes with rate $\lambda_{ap}$.	In a purely peer-to-peer environment, $\lambda_{ap}=0$. 
	\item User requests follows a Zipf-like distribution with parameter $\alpha$: 
	\begin{equation}
		p_i = \frac{\Omega}{n^\alpha}
	\end{equation}	
	%where $\Omega=\sum n^{-\alpha} \approx \frac{1-\alpha}{N^(1-\alpha)}$ 
	where $\Omega=\sum n^{-\alpha}$. We will also generalize the result to arbitrary distribution of requests.
	\item After generating a request, the user will wait for a fixed time $T$, called the \textbf{patience time}. If within time $T$, the user enters the transmission range of a wireless accessing point, or encounters another user that happens to have the requested file in its cache, then the request is a hit. Otherwise, the user needs to download the file through the 3G network and a miss is recorded. However, a long patience time is only permissible for delay tolerant networks (DTN). Even for DTN, a user is not willing to wait infinitely long. Therefore we set a hard limit on the maximal waiting time.

\begin{center}
    \begin{tabular}{ | l | p{5cm} |}
    \hline
    Symbol & Meaning \\ \hline
    $N$ & Number of files in the network \\ \hline
    $K$ & Number of files a user can cache \\ \hline
    $\lambda$ & Contact rate between two mobile users \\ \hline
    $\lambda_{ap}$ & Contact rate between a mobile user and an access point \\ \hline
    $p_n$ & Probability of request for file $n$ \\ \hline
    $T$ & Patience time	\\ \hline
    $q_n$ & Probability that a mobile user has file $n$ \\ \hline
    \end{tabular}
\end{center}

\end{enumerate}	
	\textbf{Pushing strategies}
	\vspace{3mm}
	
	We evaluate the effect of different pushing strategies for the same network. The first simple strategy is to randomly select and push files into the network, ignoring the non-uniformity of the request distribution. When a mobile user enters the transmission range of a wireless access point, the access point chooses one out of $N$ files with equal probability and transmits it to the cache of the mobile user until its cache is full. After a while, the caches of all mobile nodes become saturated and the hit rate of any file can be calculated. Therefore $P$(user has file $n$ in her cache)$= K/N$, for all $n$. 
	
	Let $\mathcal{E}_n$ be the probability that the request result for file $n$ is a miss, then
	
\begin{align}
&\mathcal{E}_n	\notag\\
=&P(\mbox{file $n$ is not in the user's cache}) P(\mbox{No encounter}	\notag\\
 &\mbox{with AP or user storing file \#n within time }T)	\notag\\
=&(1-\frac{K}{N})P(\mbox{no access point encounter within time }T)	\notag\\
&\cdot P(\mbox{no encounter with user caching file $n$})	\notag\\
=&(1-\frac{K}{N})e^{-\lambda_{ap}T}\sum_{k=0}^{\infty}P(k \mbox{ encounters})	\notag\\
 &P(\mbox{none of the } k \mbox{ users has file $n$}|k \mbox{ encounters})	\notag\\
=&(1-\frac{K}{N})e^{-\lambda_{ap}T}\sum_{k=0}^{\infty} \frac{e^{-\lambda T}(\lambda T)^k}{k!}(1-\frac{K}{N})^k	\notag\\
=&(1-\frac{K}{N})\exp\{-T[\lambda_{ap}+\lambda K/N]\}
\end{align}

For the third equality to hold, we need to assume two users encounter at most once. For a network with a very large number of users this is a realistic assumption. Since the miss rate is the same for all files, the expected miss rate for the random pushing strategy $X_{random}$ is

\begin{align}
\mathbb{E}[X_{random}]=\mathcal{E}_n=(1-\frac{K}{N})\exp\{-T[\lambda_{ap}+\lambda K/N]\}	\label{random push}
\end{align}

The observations from the above analysis are:
\begin{enumerate} 

	\item The miss ratio decreases exponentially with the patience time. Indeed, the longer a user can wait, the more likely she is to receive the requested file from either an access point or another user. 
	
	\item When $\frac{\lambda_{ap}}{\lambda}>\frac{K}{N}$, the wireless access points play the major role in file delivery. When $\frac{\lambda_{ap}}{\lambda}<\frac{K}{N}$, Peer-to-Peer transmission takes up the majority of the task. In fact, when a user generates a request for file $n$, only encounters with file $n$ holders matter to her, effectively changing the encounter rate to $\lambda K/N$.
\end{enumerate}

We now aim to improve the hit rate by considering different proactive caching strategies. With the random select and pushing strategy, we did not exploit the non-uniformity of request distribution. The overall hit rate is the average hit rate of every file, weighted by the request probability. Therefore, more efforts should be made to guarantee the delivery of the more popular files. By storing more copies of the popular files and less copies of the unpopular files, the overall hit rate could possibly improve. Suppose by strategically changing the probability of pushing each file to the mobile users, Let $q_n$ be the probability that file $n$ is stored in the cache of an arbitrary user. Then
\begin{align}
&\sum_{n=1}^N q_n=K	\\
&0\leq q_n \leq 1, n=1,\dots,N
\end{align}
Another interpretation of $q_n$ is the expected number of copies of file $n$ in a user's cache. Since a user can store $K$ files, the $q_n$'s sum up to $K$. The miss rate of request for file $n$ is then

\begin{align}
 &\mathcal{E}'_n	\notag\\
=& P(\mbox{request for file $n$ is a miss})\notag\\
=&P(\mbox{file $n$ is not stored in the user's cache}) \notag\\
 &P(\mbox{No encounter with AP or with user } \notag\\
 &\mbox{storing file $n$ within time }T)	\notag\\
=&(1-q_n)e^{-\lambda_{ap}T}\sum_{k=0}^{\infty} P(k \mbox{ encounters})	\notag\\
 &P(\mbox{none of the $k$ users has file $n$}|k \mbox{ encounters})	\notag\\
=&(1-q_n)e^{-\lambda_{ap}T}\sum_{k=0}^{\infty} \frac{e^{-\lambda T}(\lambda T)^k}{k!}(1-q_n)^k	\notag\\
=&(1-q_n)\exp\{-T(\lambda_{ap}+\lambda q_n)\}
\end{align}

The expected miss rate for the \textbf{selective pushing strategy} is
\begin{align}
\mathbb{E}[X_{select}]	=&\sum_{n=1}^N p_n \mathcal{E}'_n	\notag\\
= & p_n(1-q_n)\exp\{-T(\lambda_{ap}+\lambda q_n)\}	\notag\\
=&e^{-\lambda_{ap}T}\sum_{n=1}^N p_n(1-q_n) e^{-\lambda T q_n}	\label{selective push}
\end{align}

\vspace{3mm}

\subsection{Optimizing the Selective Pushing Strategy}
The impact of the wireless access points in (\ref{random push}) and (\ref{selective push}) is just a multiplicative factor. Therefore we ignore the effect of access points in the analyses and simulation results below. Since the sum is a convex function of $(q_1,\dots,q_n)$, the minimization problem is a convex optimization. We now cast the convex program. 

\begin{align}
\mbox{Minimize }  f(q_1,\dots,q_n) =& \sum_{n=1}^N p_n(1-q_n) e^{-\lambda T q_n}	\notag\\
\mbox{subject to }  q_n-1 \leq & 0, n=1,\dots,N	\notag\\
 -q_n \leq & 0, n=1,\dots,N	\notag\\
 \sum_{n=1}^N q_n - K =& 0	\label{convexprogram}
\end{align}

By the Karush-Kuhn-Tucker (KKT) conditions, the optimal solution $q^*$ to (\ref{convexprogram}) satisfies

\begin{align}
&\mu_n - \mu_{N+n} + \eta = p_n e^{-\lambda T q_n} (1 + \lambda T - \lambda T q_n) \notag\\
&\mbox{for } n=1,\dots, N \mbox{ (Stationarity)}	\notag\\
&0 \leq q_n \leq 1, n=1,\dots,N	\notag\\
& \sum_{n=1}^N q_n = K \mbox{ (Primal feasibility) } \notag\\
& \mu_n \geq 0, n=1,\dots,2N \mbox{ (Dual feasibility) } \notag\\
& \mu_n(q_n-1) =0, \notag\\
& \mu_{N+n} \cdot q_n = 0	\notag\\
& \mbox{for }n=1,\dots,N \mbox{ (Complementary Slackness) }
\end{align}

It remains to fix the values of $q=(q_1,\dots,q_N)$, $\mu=(\mu_1,\dots, \mu_{2N})$ and $\eta$. Observe that if $i<j$, then $q_i \geq q_j$. Otherwise since $p_i>p_j$, $f(q_1,\dots,q_n)$ could be decreased by exchanging the values of $q_i$ and $q_j$. Therefore the index set of the optimal solution $q^*$ can be divided into three parts: 

\begin{align}
&q_n = 1 \mbox{ for } n=1,\dots,N_1-1	\notag\\
&q_n\in (0,1)  \mbox{ for } n=N_1,\dots,N_2	\notag\\
&q_n = 0  \mbox{ for } n=N_2+1,\dots,N
\end{align}

The following table provides the optimal solution to (\ref{convexprogram}), except that $\eta$ is not determined:

\begin{center}
    \begin{tabular}{ | l | l | p{2cm} | p{2cm} |}
    \hline
    $n$ & $1,\dots,N_1-1$ & $N_1,\dots,N_2$ & $N_2+1,\dots, N$ \\ \hline
    $\mu_n$ & $p_n e^{-\lambda T} - \eta $ & 0 & 0 \\ \hline
    $\mu_{N+n}$ & 0 & 0 & $-p_n(1+\lambda T) + \eta$ \\ \hline
    $q_n$ & 1 & $1+\frac{1}{\lambda T} - \frac{1}{\lambda T} W( \frac{\eta e^{1+\lambda T}}{p_n})$ & 0 \\ \hline
    \end{tabular}
\end{center}

By the stationarity condition, $\eta>0$. Here $W(x)$ is the Lambert W function, the inverse function of $f(x)=xe^x$. It is single-valued and monotonically increasing on $(0,+\infty)$. Therefore $q_n$ monotonically decreases as $\eta$ increases. We propose a binary search algorithm for finding $\eta$ and $\{q_n\}$. Notice that 
\begin{align}
& \sum_n q_n = K \Rightarrow N_1 \leq K	\Rightarrow \eta \geq p_K e^{-\lambda T} \notag\\
& \sum_n q_n = K, q_n\mbox{ decreasing } \Rightarrow q_K > 0 \Rightarrow \eta < p_K(1+\lambda T)
\end{align}
Therefore, we set the boundary of search to be \newline $[p_K e^{-\lambda T}, p_K(1+\lambda T)]$

\vspace{5mm}
	\textbf{Algorithm for optimal file distribution}

	\begin{algorithmic}
	\STATE $q_n \gets 0$ for all $n$
	\STATE $\eta_{upper} \gets p_K(1+\lambda T)$
	\STATE $\eta_{lower} \gets p_K e^{-\lambda T}$
	\WHILE {$|\sum q_n - K| > \epsilon$}
		\STATE $\eta \gets (\eta_{upper}+\eta_{lower})/2$
		\FOR {$n=1$ to $N$}
			\STATE $q_n = ((1+\frac{1}{\lambda T} - \frac{1}{\lambda T} W( \frac{\eta e^{1+\lambda T}}{p_n})) \vee 0) \wedge 1$
		\ENDFOR
		\IF {$\sum q_n >K$}
			\STATE $\eta_{lower} \gets \eta$
		\ELSE
			\STATE $\eta_{upper} \gets \eta$
		\ENDIF
	\ENDWHILE
	\end{algorithmic}

\subsection{Second Model: Contact Duration-Aware}

Traditional caching schemes assume that all data requested can be transmitted within a single connection. This assumption is not likely to hold for mobile networks, where the amount data retrieved from a single contact is restricted by limited contact duration. In this scenario, a mobile user requesting a relatively large file can choose one of two strategies: (1) stop and wait when it encounters an access point or another user who happens to own the file until the transmission is complete, and (2) move on even if it can only receive a fraction of the file within the contact duration, and establish connection with other users until the transmission is complete. The analysis of the last section applies to the first strategy. In this section, we explore the performance of cooperative caching with a contact duration limitation. Due to the complexity, we only give an expression for the miss rate without solving the minimization problem. 

As in \cite{gao2009multicasting}\cite{ioannidis2010distributed}, the inter-contact time between two mobile users follows an exponential distribution. Let $N_n$ be the number of contacts between a user and other users caching file $n$ within patience time $T$, then $N_n$ follows a Poisson distribution with rate $\lambda T q_n$. Let $T_1, \dots, T_{N_n}$ represent the contact duration of each contact. They are iid random variables following the Pareto distribution (with y-axis shifted such that each time variable starts from 0): 
\begin{equation}
P[T_i \leq t] = F_{T_i}(t)= 1-(\frac{1}{t+1})^{\alpha} \mbox{ for } t\geq 0	\label{tcdf}
\end{equation}
The total contact time with users caching file $n$ can be represented as: 

\begin{equation}
T_{total,n} = T_1 + \dots + T_{N_n}
\end{equation}

where $T_{total,n}$ follows a compound Poisson distribution. Let $\varphi_{T}(t)$ be the characteristic function of $T_i$ as defined in (\ref{tcdf}), then The characteristic function of $T_{total,n}$ is

\begin{align}
\varphi_{T_{total,n}}(u) &= \mathbb{E}_{N_n} [(\varphi_{T}(u))^{N_n}] 	\notag\\
&= \sum_{k=0}^{\infty} \frac{(\lambda T q_n \varphi_{T}(u))^k e^{-\lambda T q_n }}{k!} 	\notag\\
&= e^{\lambda T q_n (1-\varphi_{T}(u))}
\end{align}

from which the cumulative distribution function of $T_{total,n}$ can be obtained. Given the file distribution $\{q_n\}$ and that each file requires a total transmission time of $t_0$, the miss rate could be calculated as follows:
\begin{align}
 \mathcal{E}_{t} =& \sum_{n=1}^N p_n P[T_{total,n}>t_0]	\notag\\
=&\sum_{n=1}^N p_n \int_0^{t_0} \frac{1}{2\pi} \int_{-\infty} ^{\infty} e^{-itu}\varphi_{T_{total,n}}(u)du
\end{align}

\section{Simulation Results}

It is difficult to precisely achieve the optimal cache allocation in practice. Wireless access points can decide on what to transmit to the users, but have no control over which user it is going to meet. We use the following \textbf{selective pushing algorithm} to achieve a near-optimal cache allocation: When a wireless access point discovers a mobile user in its transmission range, it selects file $n$ with probability $\frac{q_n}{K}$ and pushes it to the mobile user, unless the user already has the file or her cache is full. This process is repeated until all users' caches are filled. When $N \gg K$, simulation results shows that actual file distribution is very close to $(q_1,\dots,q_N)$. Figure 1 shows the case with file requests distributed according to $p_n=\beta n^{-1}$ (where $\beta$ is a normalizing constant), $N=10000$, $K=100$ and 10000 mobile users.

\begin{figure}
	\centering
		\includegraphics[width=0.45\textwidth]{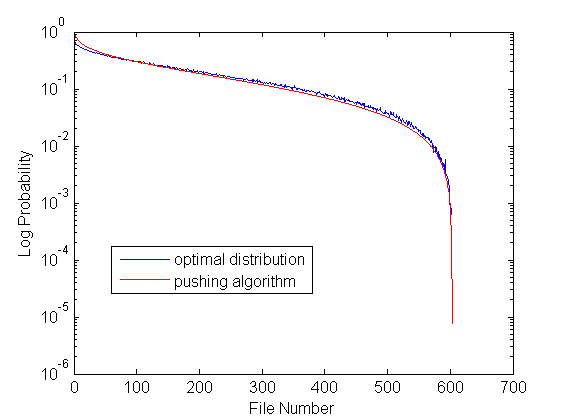}
	\caption{Comparison between the optimal caching scheme and the result of the pushing algorithm}
	\label{fig:pushingalg}
\end{figure}

Figure 2 shows the miss ratio of the three schemes versus $\lambda T$, which is the average number of encounters with other mobile users within the patience time. The value of $\lambda T$ reflects both the network mobility and the degree of delay tolerance. For comparison, we also include the traditional caching strategy, referred to as ``$K$-most popular strategy'' here, which completely ignores the opportunity of downloading files from peers and simply caches the $K$ most frequently requested files. The four curves are briefly explained:

\begin{itemize}
	\item	\textbf{Random Pushing}: Each mobile device randomly stores $K$ files in their cache.
	\item \textbf{K-most Popular}: The $K$ files with highest probability of request are cached. 
	\item \textbf{Optimal}: Files are distributed according to the solution to (8).
	\item \textbf{Pushing Algorithm}: Files are distributed as a result of the above selective pushing algorithm.
	
	\end{itemize}
 
The optimal caching scheme constantly outperforms other schemes at different values of $\lambda T$. The scheme using the above pushing algorithm performs almost identically to the optimal scheme for $\lambda T>5$, but the gap widens as $\lambda T$ decrease. When the network mobility or the delay tolerance is low, which is characterized by a small $\lambda T$, the $K$-most popular caching scheme performs very closely to the optimal caching scheme. But it is not suitable for a more dynamic or delay tolerant network.

\begin{figure}
	\centering
		\includegraphics[width=0.45\textwidth]{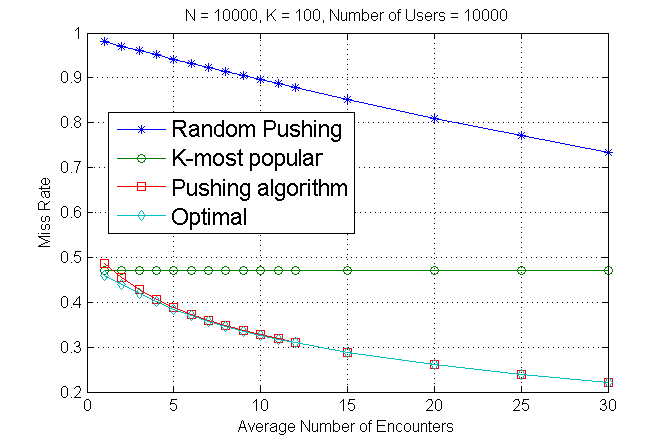}
	\caption{Miss Ratio vs $\lambda T$ for four caching schemes} 
	\label{fig:k100n1000user10000}
\end{figure}

\section{Conclusions and Future Work}

In this work, we studied the effect of cooperative caching schemes on a delay tolerant mobile network, where content requests are generated subject to a probability distribution. Our main contribution is to identify the optimal file allocation to each mobile user's cache in a homogeneous environment where all users share the same mobility or centrality statistic. We showed that in a network with higher user mobility or delay tolerance, the performance of our scehme shows significant improvement than the traditional caching scheme. We proposed a simple algorithm to achieve the near-optimal file allocation. Our work is the first to take the content popularity into consideration. 

The current work has the potential to be extended in a number of ways. A model based on social network, where each user has different mobility and centrality characteristics may be of interest. In such a model some users would have a higher encounter rate with other users and hence play a more important role in peer-to-peer transmission. If we drop the assumption of homogeniety, i.e. allow mobile users to have different caching capacity, it may be interesting to study whether the optimal caching strategy is essentially the same. It may also be interesting to consider dynamic content evolution, where popularity of a content is a function of time, rather than static. This would require us to design a scheme to replace or reshuffle users' cache content over time.

\bibliographystyle{plain}
\bibliography{Aug1}

\end{document}